\documentclass[prd,aps,showpacs,nofootinbib,twocolumn,superscriptaddress,
amssymb]{revtex4}

\usepackage{graphicx}
\usepackage[english]{babel}
\usepackage{amsmath}
\usepackage{amssymb}
\usepackage{amsfonts}
\usepackage{latexsym}
\usepackage{color,amsxtra}
\usepackage{epsf}
\usepackage{enumerate}
\usepackage{hhline}
\usepackage{array}
\usepackage{tabularx}
\usepackage{subfigure}
\usepackage{fancyhdr}
\usepackage{mathrsfs}
\newcommand{\be}{\begin{equation}}
\newcommand{\ee}{\end{equation}}
\newcommand{\bea}{\begin{eqnarray}}
\newcommand{\eea}{\end{eqnarray}}
\newcommand{\beaa}{\begin{eqnarray*}}
\newcommand{\eeaa}{\end{eqnarray*}}

\newcommand{\Eqn}[1]{&\hspace{-0.2em}#1\hspace{-0.2em}&}
\def\Vec#1{\mbox{\boldmath $#1$}}

\def\be{\begin{equation}}
\def\ee{\end{equation}}
\def\bea{\begin{eqnarray}}
\def\eea{\end{eqnarray}}

\begin{document}

\title{Inflationary cosmology in unimodular $F(T)$ gravity}

\author{Kazuharu Bamba}
\email{bamba@sss.fukushima-u.ac.jp}
\affiliation{Division of Human Support System, Faculty of Symbiotic Systems Science, 
Fukushima 
University, 
Fukushima 960-1296, Japan}

\author{Sergei D. Odintsov}
\email{odintsov@ieec.uab.es}
\affiliation{Institut de Ciencies de lEspai (IEEC-CSIC), 
Campus UAB, Carrer de Can Magrans, s/n 
08193 Cerdanyola del Valles, Barcelona, Spain}
\affiliation{Instituci\'{o} Catalana de Recerca i Estudis Avan\c{c}ats
(ICREA), Passeig Llu\'{i}s Companys, 23 08010 Barcelona, Spain}

\author{Emmanuel N. Saridakis}
\email{Emmanuel\_Saridakis@baylor.edu}
\affiliation{CASPER, Physics Department, Baylor University, Waco, TX 76798-7310, USA}
\affiliation{Instituto de F\'{\i}sica, Pontificia Universidad de Cat\'olica de 
Valpara\'{\i}so, 
Casilla 4950, Valpara\'{\i}so, Chile}


\begin{abstract} 
We investigate the inflationary realization in the context of unimodular $F(T)$ gravity, 
which is based on the $F(T)$ modification of  teleparallel gravity, in which one imposes 
the unimodular condition through the use of Lagrange multipliers. We develop 
the general reconstruction procedure of the $F(T)$ form  that can give rise to a given 
scale-factor evolution, and then we apply it in the inflationary regime. We extract 
the Hubble slow-roll parameters that allow us to calculate 
various inflation-related observables, such as the scalar spectral index and its running, 
the tensor-to-scalar ratio, and the tensor spectral index. Then, we examine the 
particular cases of de Sitter and power-law inflation, of Starobinsky inflation, as well 
as inflation in a specific model of unimodular $F(T)$ gravity. As we show, in all cases 
the predictions of our scenarios are in a very good agreement with Planck observational 
data. Finally, inflation in unimodular $F(T)$ gravity has the additional 
advantage that it always allows for a graceful exit for specific regions of the model 
parameters.
\end{abstract}



\maketitle



\section{Introduction}

According to the Standard Model of Cosmology, supported by a large amount of 
observational 
data, during its evolution the universe has experienced two phases of accelerated 
expansion, the inflationary one at early times and the current one at late times 
\cite{SN,Planck:2015xua, Array:2015xqh, Hinshaw:2012aka,LSS,Jain:2003tba}. 
In order to explain these accelerated phases one should introduce small deviations to 
the standard physics paradigm, and there are two main directions that one can follow. The 
first is to maintain general relativity as the gravitational theory and alter the content 
of the universe by introducing novel, exotic forms of matter fields, either as inflaton 
field(s) \cite{Inflation} or as dark energy sector \cite{Peebles:2002gy,Cai:2009zp}. 
The second way is to modify the gravitational sector, constructing a theory that 
possesses 
general relativity as a particular limit, but with additional degrees of freedom that can 
drive an accelerating expansion \cite{R-D-M}. 

Although most of the works in modified gravity start from the usual gravitational 
description based on curvature and modify the Einstein-Hilbert action, with the 
simplest example being the $F(R)$ scenario \cite{F-R}, one could equally well construct 
gravitational modifications starting from the torsion-based description of gravity. In 
particular, one could start from the Teleparallel Equivalent of General 
Relativity (TEGR) \cite{ein28,Hayashi79,Pereira.book,Maluf:2013gaa}, in which 
the gravitational Lagrangian is the torsion scalar $T$, and build various extensions, 
with 
the simplest example being the $F(T)$ theory \cite{Bengochea:2008gz,Linder:2010py} (for a 
review see \cite{Cai:2015emx}). Interestingly enough, although TEGR is completely 
equivalent with general relativity at the level of equations, $F(T)$ does not coincide 
with $F(R)$ gravity, and thus its cosmological application has led to novel features, 
either at inflationary stage \cite{F-T-Inf} or at the late-time accelerated epoch 
\cite{Chen:2010va,Geng:2011aj}. 

On the other hand, unimodular gravity \cite{U-G} is an interesting gravitational theory, 
which can be considered as a specific case of general relativity. In particular, while in 
standard general relativity the origin of the cosmological constant is not well 
understood 
\cite{C-C}, in unimodular gravity it arises as the trace-free part of the gravitational 
field equations, as long as the determinant of the metric is fixed to a number or a 
function. The great theoretical advantage of this procedure is that since the trace-free 
part of the field equations is not related to the vacuum expectation value of any matter 
field, one can fix its value without facing the cosmological constant problem. Hence, one 
can use unimodular gravity in order to describe the inflationary regime 
\cite{Cho:2014taa} 
and the late-time cosmic acceleration \cite{UG-LC}. Additionally, one can start from 
unimodular gravity in order to construct extensions, such as unimodular $F(R)$ gravity 
\cite{Nojiri:2015sfd} and unimodular $F(T)$ gravity \cite{Nassur:2016yhc}, which prove to 
have interesting cosmological implications.

In the present work we are interesting in investigating inflationary cosmology in the 
framework of unimodular $F(T)$ gravity. Specifically, we study the Lagrange multiplier 
method to represent the action of unimodular $F(T)$ gravity and we build its  
reconstruction procedure. In addition, we extract the 
observables of the inflationary regime, 
namely the spectral index of the curvature perturbations and its running, the tensor 
spectral index, and the tensor-to-scalar ratio, 
 showing that they 
are in very satisfactory agreement with observations. Finally, we study the instability 
of the de Sitter solution, by investigating its perturbations, showing that a graceful 
exit can always be realized for specific parametric regions, which is an advantage of 
inflation in unimodular $F(T)$ gravity.

The plan of the work is the following. 
In Sec. \ref{unimodulars} we formulate unimodular $F(T)$ gravity using Lagrange 
multipliers and we present the reconstruction procedure. In Sec. \ref{Inflationarys} we 
investigate the inflationary realization in the context of unimodular $F(T)$ gravity, in 
the case of de Sitter, power-law and Starobinsky inflation, as well as for a specific 
$F(T)$ form, extracting the inflationary observables and comparing them with the Planck 
data. In Sec. \ref{grexit} we discuss the graceful exit from inflation in the scenario at 
hand. Finally, Sec. \ref{Conclusions} is devoted to the conclusions.

\section{Unimodular $F(T)$ gravity}
\label{unimodulars}

In this section we will present unimodular $F(T)$ gravity and then we will formulate it 
using Lagrange multipliers. Finally, we will provide the method for reconstructing 
unimodular $F(T)$ gravity in the general case.

\subsection{Teleparallel and $F(T)$ gravity} 

In teleparallel formulation of gravitation one uses the vierbeins $e^\mu_A$ as dynamical 
variables, which at each point $x^{\mu}$ of a generic manifold form an orthonormal base 
for the tangent space. The metric is then given as 
\begin{equation}
g_{\mu\nu}=\eta_{A B} e^A_\mu 
e^B_\nu,
\label{metricvierbein}
\end{equation}
 where greek indice span the coordinate space while latin indices span the 
tangent space). Furthermore, one uses the curvatureless
Weitzenb{\"{o}}ck connection  \cite{Pereira.book} 
$\overset{\mathbf{w}}{\Gamma}^\lambda_{\nu\mu}\equiv e^\lambda_A\:\partial_\mu e^A_\nu$, 
instead of the standard torsionless Levi-Civita one, and thus the gravitational field is 
described not by the curvature tensor but by the torsion one, which reads as
\begin{equation}
T^\rho_{\verb| |\mu\nu} \equiv e^\rho_A
\left( \partial_\mu e^A_\nu - \partial_\nu e^A_\mu \right).
\end{equation}
The Lagrangian of such a theory is just the torsion scalar $T$, which is constructed by
contractions of the torsion tensor, namely \cite{Maluf:2013gaa}
\begin{equation}
\label{Tdefscalar}
T\equiv\frac{1}{4}
T^{\rho \mu \nu}
T_{\rho \mu \nu}
+\frac{1}{2}T^{\rho \mu \nu }T_{\nu \mu\rho}
-T_{\rho \mu}{}^{\rho }T^{\nu\mu}{}_{\nu}.
\end{equation}
Since in $F(R)$ gravity one extends the Einstein-Hilbert Lagrangian, namely the Ricci 
scalar $R$ to an arbitrary function $F(R)$, one can follow a similar procedure 
in the context of teleparallel gravity, i.e. generalize $T$ to $F(T)$ obtaining the 
$F(T)$ gravitational modification \cite{Bengochea:2008gz,Linder:2010py,Cai:2015emx}:
\begin{equation}
  S= \int d^4x e \left[
\frac{F(T)}{2{\kappa}^2}
\right],
\label{actiontotfT}
\end{equation}
 where
$e= \det \left(e^A_\mu \right)=\sqrt{-g}$ and $\kappa^2 =8\pi G=M_p^{-1}$ is the 
gravitational constant, with $M_p$ the Planck mass.

The equations of motion for $F(T)$ gravity arise by variation of the total action
$S+ S_{\mathrm{M}}$, where $ S_{\mathrm{M}}$ is the matter action, in terms of the
vierbeins, and they write as
\begin{eqnarray}
\label{eomsgeneral}
&&\!\!\!\!\!\!\!e^{-1}\partial_{\mu}(ee_A^{\rho}S_{\rho}{}^{\mu\nu})\frac{d
F(T)}{dT} 
 +
e_A^{\rho}S_{\rho}{}^{\mu\nu}\partial_{\mu}({T})\frac{d^2 F(T)}{dT^2} \ \ \ \ \  \  
\ \
\ \ \nonumber\\
&&\!  
-\frac{d
F(T)}{dT}  e_{A}^{\lambda}T^{\rho}{}_{\mu\lambda}S_{\rho}{}^{\nu\mu}+\frac{1}{4} e_ { A
} ^ {\nu}F(T) = \frac{{\kappa}^2}{2} e_{A}^{\rho}\,{T^{(\mathrm{M})}}_\rho^{\verb| |\nu},
\end{eqnarray}
where the ``super-potential'' tensor $S_\rho^{\verb| |\mu\nu} = \frac{1}{2}
\left(K^{\mu\nu}_{\verb|  |\rho}+\delta^\mu_\rho
  T^{\alpha \nu}_{\verb|  |\alpha}-\delta^\nu_\rho
 T^{\alpha \mu}_{\verb|  |\alpha}\right)$ is defined in terms of the co-torsion tensor
$K^{\mu\nu}_{\verb| |\rho}=-\frac{1}{2}\left(T^{\mu\nu}_{\verb|  |\rho} - T^{\nu
\mu}_{\verb|  |\rho} - T_\rho^{\verb| |\mu\nu}\right)$. Additionally, ${
T^{({\mathrm{M}})}}_\rho^{\verb| |\nu}$ denotes the energy-momentum tensor 
corresponding to $S_{\mathrm{M}}$. We mention that in the case where $F(T)=T$ one obtains
teleparallel equivalent of general relativity, in which case equations 
(\ref{eomsgeneral}) coincide with the field equations of the latter.

\subsection{Unimodular conditions}

Let us now present briefly the basic idea of unimodular gravity. In such a construction
the determinant $g$ of the metric  $g^{\mu\nu}$ is imposed to be a constant value, 
namely the metric components are constrained in order for $\sqrt{-g}$ to be fixed. 
Without 
loss of generality, one can set $\sqrt{-g} = 1$ \cite{Nojiri:2015sfd}. In the case of 
cosmological applications one considers a flat 
Friedmann-Lema\^{i}tre-Robertson-Walker (FLRW) space-time 
with metric 
\begin{equation} 
ds^2 = dt^2 - a^2(t) 
\left[ \left(dx^1\right)^2 + \left(dx^2\right)^2 + 
\left(dx^3 \right)^2 \right]\,, 
\label{eq:2.5}
\end{equation} 
which arises from the vierbein
\begin{equation} 
e_{\mu}^A={\rm diag}(1,a(t),a(t),a(t)) 
\label{eq:2.5vierbein},
\end{equation}
where $a(t)$ is the scale factor and $t$ is the cosmic time. Introducing a new time 
variable $\tau$  through
\begin{equation}  
d\tau \equiv a^3 (t) dt,
\label{dtdatu}
\end{equation}
the FLRW metric is rewritten as 
\begin{equation}  
ds^2 = a^{-6} (\tau) d\tau^2 
- a^2(\tau) \left[ \left(dx^1\right)^2 + \left(dx^2\right)^2 + 
\left(dx^3 \right)^2 \right]\,, 
\label{eq:2.6}
\end{equation} 
with $a(\tau) \equiv a(t(\tau))$. 
In this case we have $g_{\mu \nu}= \mathrm{diag} (a^{-6}(\tau), -a^2(\tau), 
-a^2(\tau), -
a^2(\tau))$ and the vierbein components are given by 
\begin{equation} 
e^A_\mu = \mathrm{diag}(a^{-3}(\tau), a(\tau), a(\tau), a(\tau))
\label{eq:2.6vierbein2}.
\end{equation}
 Hence, we can easily verify the satisfaction of the unimodular gravity  constraint, 
namely that  $|e|= \det \left(e^A_\mu \right)=\sqrt{-g} =1$.


\subsection{Lagrange multiplier formulation of unimodular $F(T)$ gravity}

In this subsection we will present the Lagrange multiplier formulation of unimodular 
$F(T)$ gravity, following the corresponding procedure developed for $F(R)$ gravity in 
\cite{Nojiri:2015sfd}. In particular, we will use the Lagrange multiplier 
method \cite{L-M-M} in the framework of $F(T)$ gravity in order to ensure that the 
unimodular condition is satisfied. Introducing the Lagrange multiplier $\lambda$, 
the action of unimodular $F(T)$ gravity with matter can be written 
as \cite{Nassur:2016yhc}
\begin{equation}
S = \int d^4x \left\{ |e| \left[ \frac{F(T)}{2{\kappa}^2} 
-\lambda \right] + \lambda \right\}
+{S}_{\mathrm{M}} \,, 
\label{eq:2.15}
\end{equation}
and in the following we set $2{\kappa}^2 = 1$ for simplicity. 

By varying the action in Eq.~(\ref{eq:2.15}) with respect to 
the vierbein we acquire \cite{Nassur:2016yhc} 
\begin{eqnarray}
\label{eq:2.16}
&&\!\!\!\!\!\!\!\!\!\!\!\!\!\!\!\!\!
e^{-1}\partial_{\mu}(ee_A^{\rho}S_{\rho}{}^{\mu\nu})\frac{d
F(T)}{dT} 
 +
e_A^{\rho}S_{\rho}{}^{\mu\nu}\partial_{\mu}({T})\frac{d^2 F(T)}{dT^2}  
\nonumber\\
&&\!\!\!\!\!\!\!\!\!\!\!\!\!\!\!\!\!
-\frac{d
F(T)}{dT}  e_{A}^{\lambda}T^{\rho}{}_{\mu\lambda}S_{\rho}{}^{\nu\mu}\!+\!\frac{1}{4} e_ { 
A
} ^ {\nu}\left[F(T)\! - \!\lambda \right] \!=\! \frac{1}{4} 
e_{A}^{\rho}\,{T^{(\mathrm{M})}}_\rho^{\verb| |\nu}\!.
\end{eqnarray}
In the following we consider the matter energy-momentum tensor 
${T^{(\mathrm{M})}}_\rho^{\verb| |\nu}$ to correspond to a perfect fluid, namely
  ${T^{(\mathrm{M})}}_\rho^{\verb| |\nu} = 
\mathrm{diag} (\rho_{\mathrm{M}}, -P_{\mathrm{M}}, -P_{\mathrm{M}}, -P_{\mathrm{M}})$, 
where $\rho_{\mathrm{M}}$ and $P_{\mathrm{M}}$ are the energy density 
and pressure respectively. 

In the case of FRLW geometry of (\ref{eq:2.5vierbein}) or (\ref{eq:2.6vierbein2}), the 
torsion scalar $T$ defined in (\ref{Tdefscalar}) becomes
\begin{eqnarray}
T=-6 H(t)^2=-6 a^6 (\tau) \mathcal{H}(\tau)^2,
\label{Tscalartau}
\end{eqnarray}
where 
$\mathcal{H}(\tau) \equiv \frac{1}{a(\tau)}\frac{d a(\tau)}{d\tau}$ is the new function 
that plays the role of the 
Hubble parameter $H(t) \equiv \dot{a}(t)/a(t)$ (with dots denoting derivatives with 
respect to $t$). Hence, in this case the general field equations 
(\ref{eq:2.16}) give rise to the two Friedmann equations as
\begin{equation}
12a^6(\tau)\mathcal{H}^2 \frac{d F(T)}{dT} 
+ \left[ F(T) - \lambda \right] -\rho_{\mathrm{M}} =0\,, 
\label{eq:2.17}
\end{equation}
\begin{eqnarray}
&&\!\!\!\!\!\!\!\!\!\!\!\!\!\!\!\!\!\!\!\!\!
-48a^{12}(\tau)\mathcal{H}^2 
\left(3\mathcal{H}^2 + \frac{d\mathcal{H}}{d\tau}\right) 
\frac{d^2 F(T)}{dT^2}+ \left[ F(T) - \lambda \right] 
\nonumber\\
&&\ \ \ 
+4a^6(\tau) \left( 6\mathcal{H}^2 
+ \frac{d\mathcal{H}}{d\tau}\right)\frac{d F(T)}{dT} 
+ P_{\mathrm{M}} =0 \,.
\label{eq:2.18} 
\end{eqnarray} 
Eliminating the term $\left( F(T) - \lambda \right)$ between (\ref{eq:2.17}) and 
(\ref{eq:2.18}), and using the relation (\ref{Tscalartau})
we acquire
\begin{equation}
\!
4a^6(\tau) \!
\left(3\mathcal{H}^2 + \frac{d\mathcal{H}}{d\tau}\right) \!
\left[2T\frac{d^2 F(T)}{dT^2}\!+\!\frac{d F(T)}{dT}\right]
+ \rho_{\mathrm{M}} + P_{\mathrm{M}} = 0\,. 
\label{eq:2.20}
\end{equation}
Finally, the system of equations closes by considering the  continuity equation for the 
matter fluid, namely 
\begin{equation}
\frac{d\rho_{\mathrm{M}}}{d\tau} + 3\mathcal{H} 
\left(\rho_{\mathrm{M}} + P_{\mathrm{M}}\right) = 0\,. 
\label{eq:2.19}
\end{equation}

\subsection{Reconstruction of unimodular $F(T)$ gravity}
\label{lagmultform}

In this subsection we will present the method of reconstructing the $F(T)$ form that 
generates a given scale-factor evolution. We start by differentiating (\ref{Tscalartau})
in order to obtain the useful relation 
\begin{equation}
\frac{dT}{d\tau} = -12a^6(\tau)\mathcal{H} 
\left(3\mathcal{H}^2 + d\mathcal{H}/d\tau \right),
\label{dTdtau}
\end{equation}
and inserting it into Eq. (\ref{eq:2.20}) we acquire
\begin{equation}
-\frac{1}{3\mathcal{H}} 
\left[2T\frac{d}{d\tau} 
\left(\frac{d F(T)}{dT}\right) 
+ \frac{dF(T)}{d\tau} 
\right] 
+ \rho_{\mathrm{M}} + P_{\mathrm{M}} = 0\,. 
\label{eq:2.23}
\end{equation}

Let us now consider a specific scale factor. For simplicity we choose the general 
power-law form
 \begin{equation}
 a(t) = a_* \left( \frac{t}{t_*} \right)^p,
 \label{apowergener}
 \end{equation}
 where $a_*$ is a value of $a$ at time $t_*$ and 
$p $ is a constant, but the reconstruction procedure can be applied in a general 
$a(t)$ too. In this case relation (\ref{dtdatu}) leads to
\begin{eqnarray} 
\tau \Eqn{=} \frac{a_*^3\,t_* }{3p+1} \left(\frac{t}{t_*}\right)^{3p+1}\,, 
\label{tautrelation}
\end{eqnarray} 
and in terms of the new time variable $\tau$ the above scale factor reads as  
\begin{eqnarray} 
a(\tau) \Eqn{=} \left(\frac{\tau}{\tau_*}\right)^q \,, 
\label{eq:2.1200}
\end{eqnarray} 
with 
\begin{eqnarray} 
q \Eqn{\equiv} \frac{p}{3p+1} \,,
\nonumber\\ 
\tau_* \Eqn{\equiv} \frac{t_*}{a_*^{1/p} \left(3p+1\right)} \,. 
\label{eq:2.14}
\end{eqnarray} 
In this case $H=p/t$ and   $\mathcal{H} = q/\tau$, and hence (\ref{Tscalartau}) leads to
\begin{equation}
T= -6\left(\frac{q}{\tau_*}\right)^2 
\left(\frac{\tau}{\tau_*}\right)^{2\left(3q-1\right)}.
\label{Ttauaux}
\end{equation}
 Thus, we can easily find that
$d/dT = -\left\{\tau_*^3/\left[12q^2\left(3q-1\right)\right]\right\} 
\left(\tau/\tau_*\right)^{-3\left(2q-1\right)} d/d\tau$. 
 Additionally, if the matter perfect fluid has a constant equation-of-state parameter 
$w=P_{\mathrm{M}}/\rho_{\mathrm{M}}$, then the continuity equation (\ref{eq:2.19}) gives
$\rho_{\mathrm{M}} = \rho_{\mathrm{M}*} \left(\tau/\tau_*\right)^{-3q\left(1+w\right)}$, 
where $\rho_{\mathrm{M}*}$ is the value of $\rho_{\mathrm{M}}$ 
at $\tau = \tau_*$. Inserting these into Eq. (\ref{eq:2.23})  we obtain 
\begin{eqnarray} 
&&\!\!\!\!\!\!\!\!\!\!\!\!\!\!\!\!\!\!\!\!\!
\frac{d^2F(\tau)}{d\tau^2} + \frac{\left(2-3q\right)}{\tau} \frac{dF(\tau)}{d\tau} 
\nonumber\\
&&
-\frac{3q\left(
3q-1\right)\left(1+w\right)\rho_{\mathrm{M}*}}{\tau_*^{-3q\left(1+w\right)}} 
\tau^{-3q\left(1+w\right)-2} = 0\,,
\label{eq:2.24}
\end{eqnarray}
which is a differential equation in terms of $\tau$. The general solution of  
(\ref{eq:2.24}) reads as
\begin{equation}  
\!\!
F(\tau) = c_1 \tau^{3q-1} 
- 
\frac{\left(3q-1\right)\rho_{\mathrm{M}*}}{\left[3q\left(2+w\right)-1\right] 
}\left(\frac{ \tau}{\tau_*}\right)^{-3q\left(1+w\right)} +c_2 , 
\label{eq:2.25}
\end{equation}
where $c_1$ and $c_2$ are integration constants. Therefore, using (\ref{Ttauaux}) we can 
express this solution in terms of $T$ as
\begin{eqnarray}  
&&\!\!\!\!\!\!\!\!\!\!\!\!\!\!\!\!\!\!\!
F(T) = c_1 \frac{\tau_*^{3q}}{ \sqrt{6}q} \sqrt{-T} +c_2\nonumber\\
&&\!\!\!\!\!\!\!\!\!\!\!\!\!\!
- \frac{\left(3q-1\right)\left(6q^2\right)
^{\frac{3q\left(1+w\right)}{2\left(3q-1\right)}} 
\tau_*^{-\frac{3q\left(1+w\right)}{3q-1}} 
\rho_{\mathrm{M}*}}{\left[3q\left(2+w\right)-1\right]}  
(-T)^{-\frac{3q\left(1+w\right)}{2\left(3q-1\right)}}  .
\label{eq:2.26}
\end{eqnarray}
Hence, we have reconstructed the $F(T)$ form that generates the power-law scale factor 
evolution (\ref{eq:2.1200}). Finally, for completeness we give the expression for the 
Lagrange multiplier too. In particular, inserting (\ref{eq:2.26})  into 
Eq.~(\ref{eq:2.17}) we find that  
\begin{equation}  
\lambda (\tau) = -2 \rho_{\mathrm{M}*} 
\left(\frac{\tau}{\tau_*}\right)^{-3q\left(1+w\right)} + c_2 
\,,
\label{eq:2.27}
\end{equation}
which using (\ref{Ttauaux}) leads to 
\begin{equation}  
\lambda (T) = 
-2\rho_{\mathrm{M}*} \left(6q^2\right)^{\frac{3q\left(1+w\right)}{2\left(3q-1\right)}} 
\tau_*^{-\frac{3q\left(1+w\right)}{3q-1}} 
(-T)^{-\frac{3q\left(1+w\right)}{2\left(3q-1\right)}} +c_2 \,.
\label{eq:2.28}
\end{equation}
Lastly, note that in the vacuum case, i.e. when $\rho_{\mathrm{M}} =0$ and 
$P_{\mathrm{M}} = 0$, we find that
\begin{equation}
F(T) = c_1 \sqrt{-T} + c_2 \,, 
\label{eq:2.21}
\end{equation}
while
\begin{equation}
\lambda = c_2 \,. 
\label{eq:2.22}
\end{equation}
Thus, the Lagrange multiplier becomes constant and the $F(T)$ form can be reconstructed 
without using the expression of the scale factor. However, we mention that in this simple 
case the theory becomes trivial, since the Lagrangian, in the absence of the constant 
term, becomes a total derivative \cite{Basilakos:2013rua}.

\section{Inflationary Cosmology}
\label{Inflationarys}

In the previous section we presented unimodular $F(T)$ gravity and we analyzed the 
procedure with which one can reconstruct the specific $F(T)$ form that can generate a 
given scale-factor evolution. Hence, in the present section we will apply these in 
inflationary cosmology, in order to investigate inflation realization in unimodular 
$F(T)$ 
gravity. Additionally, we will extract various inflation-related observables, such as 
the scalar and tensor spectral indices, the running spectral index, and the 
tensor-to-scalar ratio, and we will compare them with observational data.

\subsection{Slow-roll parameters and inflationary observables} 

In every inflationary scenario one needs to calculate the values of various 
inflation-related observables, such as the scalar spectral index of the curvature 
perturbations $n_\mathrm{s}$, the running $\alpha_\mathrm{s} \equiv d n_\mathrm{s}/d 
\ln k$ of the spectral index 
$n_\mathrm{s}$, where $k$ is the absolute value of the wave number $\Vec{k}$,
the tensor spectral 
index $n_\mathrm{T}$ and the tensor-to-scalar ratio $r$, since these quantities are 
determined very accurately by observational data, and thus confrontation can constrain of 
exclude the scenarios at hand. In principle, the calculation of the above observables 
requires a detailed and lengthy perturbation analysis. However, one can bypass this 
procedure by transforming the given scenario to the Einstein frame, where all the 
inflation information is encoded in the (effective) scalar potential $V(\phi)$, defining 
the slow-roll parameters $\epsilon$, $\eta$ and $\xi$ in terms of this potential and its 
derivatives as \cite{IM-PDP,Martin:2013tda}
\begin{eqnarray}
\label{epsV}
\epsilon&\equiv&\frac{M_p^2}{2}\left(\frac{1}{V}\frac{ dV}{ d\phi}\right)^2\, 
,~~\,
\,\\
\eta &\equiv& \frac{M_p^2}{V}\frac{ d^2V}{ d\phi^2}\,
,
\label{etaV}\\
\xi^2&\equiv&\frac{M_p^4}{V^2}\frac{ dV}{ d\phi}\frac{
 d^3V}{d\phi^3}\, ,
\label{xiV}
\end{eqnarray}
(inflation ends when $\epsilon=1$),
and then using the approximate expressions for the observables in terms of 
these potential-related slow-roll parameters \cite{Martin:2013tda}:
\begin{eqnarray}
 r &\approx&16\epsilon ,
 \label{eps111}\\
 n_\mathrm{s} &\approx& 1-6\epsilon+2\eta  ,
\label{eps2222} \\
\alpha_\mathrm{s} &\approx& 16\epsilon\eta-24\epsilon^2-2\xi^2 , 
\label{eps333}
\\
 n_\mathrm{T} &\approx& -2\epsilon .
\label{eps444}
\end{eqnarray}

However, the above procedure cannot be applied in modified gravity scenarios where the 
conformal transformation to the Einstein frame is absent, since in this case one cannot 
define a scalar potential and then the potential-related slow-roll parameters. In such 
scenarios one should instead introduce the ``Hubble slow-roll'' parameters $\epsilon_n$ 
(with $n$ positive integer), defined as \cite{Martin:2013tda}
\begin{eqnarray}
\epsilon_{n+1}\equiv \frac{d\ln |\epsilon_n|}{dN},
\end{eqnarray}
with $\epsilon_0\equiv H_{ini}/H$ and $N\equiv\ln(a/a_{ini})$  the e-folding number, 
and where $a_{ini}$ is the scale factor at the beginning of inflation and $H_{ini}$ the 
corresponding Hubble parameter (inflation ends when $\epsilon_1=1$). In terms of the 
first three $\epsilon_n$, which are straightforwardly extracted to be 
\begin{eqnarray}
\label{epsVbb}
&&\!\!\!\!\!\!\!\!\!\!\!\!\!\!
\epsilon_1\equiv-\frac{\dot{H}}{H^2}, 
\\
&&\!\!\!\!\!\!\!\!\!\!\!\!\!\!
\epsilon_2 \equiv  \frac{\ddot{H}}{H\dot{H}}-\frac{2\dot{H}}{H^2},
\label{etaVbb}\\
&&\!\!\!\!\!\!\!\!\!\!\!\!\!\!
\epsilon_3 \equiv
\left(\ddot{H}H-2\dot{H}^2\right)^{-1}
\nonumber\\
&&\!\!\!\!
\cdot\!\left[\frac{H\dot{H}\dddot{H}-\ddot{H}(\dot{H}
^2+H\ddot{H}) } { H\dot { H } }-\frac{2\dot{H}}{H^2}(H\ddot{H}-2\dot{H}^2)
\right]\!,
\label{xiVbb}
\end{eqnarray}
the inflationary observables write as \cite{Martin:2013tda}
 \begin{eqnarray}
 r &\approx&16\epsilon_1 ,
 \label{eps111bb}\\
 n_\mathrm{s} &\approx& 1-2\epsilon_1-2\epsilon_2  ,
\label{eps2222bb} \\
\alpha_\mathrm{s} &\approx& -2 \epsilon_1\epsilon_2-\epsilon_2\epsilon_3  ,
\label{eps333bb}\\
 n_\mathrm{T} &\approx& -2\epsilon_1  .
\label{eps444bb}
\end{eqnarray}
Obviously, in cases where both the potential-related slow-roll parameters and the Hubble 
slow-roll parameters can be defined, the final expressions for the observables $r$, 
$ n_\mathrm{s}$, $\alpha_\mathrm{s}$ and $ n_\mathrm{T}$ coincide.

In the present work we are interested in investigating inflationary cosmology in the 
framework of unimodular $F(T)$ gravity. Similarly to usual  $F(T)$ gravity, and in 
contrast with $F(R)$ gravity, in this case there is not a conformal transformation to the 
Einstein frame \cite{Cai:2015emx,Yang:2010ji}, where one could find the effective scalar 
potential and then use it in order to calculate  the potential-related slow-roll 
parameters. Hence, in order to calculate the inflationary observables we must use the 
Hubble slow-roll parameters (\ref{epsVbb})-(\ref{xiVbb}), and then insert them 
into expressions (\ref{eps111bb})-(\ref{eps444bb}). 

Finally, let us stress here that the above procedure pre-assumes that there is only one 
extra (comparing to general relativity) degree of freedom that drives inflation, and which 
will affect the scalar and tensor perturbations. However, in general, $F(T)$ gravity 
possesses three extra degrees of freedom, corresponding to one massive vector field or to 
one massless
vector field and one scalar  \cite{Li:2011rn}. Nevertheless, as it was shown in 
\cite{Li:2011wu}, in the case of FLRW geometry, and for viable $F(T)$ 
scenarios, only one extra degree of freedom arises at the background and linear 
perturbation levels. Hence, as a first approximation, the observables can be calculated 
with the aforementioned procedure, through the use of Hubble slow-roll parameters.

\subsection{de Sitter and power-law inflation}

Let us first provide the kinematical expressions for inflation realization. Without loss 
of generality we focus on two basic inflationary scale-factor evolutions, namely the de 
Sitter and the power-law ones. The investigation of arbitrary evolutions is 
straightforward. 

For the de Sitter inflation the scale factor has the well-known exponential form 
\begin{eqnarray} 
a(t) = e^{ H_\mathrm{inf} t },
\label{deSittat}
\end{eqnarray}
where 
$H_\mathrm{inf}$ is the constant Hubble parameter at the inflationary stage. 
In this case,  following the procedure of the previous section, and in particular  
relations (\ref{dtdatu}) and (\ref{eq:2.6}), we find
\begin{eqnarray} 
\tau \Eqn{=} \frac{1}{3H_\mathrm{inf}} 
e ^{ 3H_\mathrm{inf} t}\,, 
\label{eq:2.7}\\
a (\tau) \Eqn{=} \left( H_\mathrm{inf} \tau \right)^{1/3}\,.
\label{eq:2.8}
\end{eqnarray} 
Similarly, for the power-law inflation the scale factor has the form (\ref{apowergener}), 
namely 
 \begin{equation}
 a(t) = a_* \left( \frac{t}{t_*} \right)^p,
 \label{apowergener2}
 \end{equation}
 where $a_*$ is the value of $a$ at time $t_*$ and 
$p$ is a constant. In this case, and as was analyzed in the previous section, in 
expressions (\ref{eq:2.1200}) and (\ref{eq:2.14}), we obtain
\begin{eqnarray} 
a(\tau) \Eqn{=} \left(\frac{\tau}{\tau_*}\right)^q \,, 
\label{eq:2.1200bb}
\end{eqnarray} 
with 
$q = p/(3p+1) $ and $
\tau_* = t_*/[a_*^{1/p} (3p+1)] $. Note that in the limit $p \to \infty$ and $t_* \to 
\infty$ with  $ p/t_*=H_*=const.$, the power-law expansion (\ref{apowergener2}) gives 
the de Sitter expansion (\ref{deSittat}), with $H_\mathrm{inf}=3 H_*$. Hence, we can 
study both cases in a simultaneous way, using (\ref{eq:2.1200bb}), and thus the 
unimodular metric (\ref{eq:2.6}) becomes
\begin{equation}  \!
ds^2 = \left( \frac{\tau}{\tau_*} \right)^{-6q}\!\! d\tau^2 
- \left( \frac{\tau}{\tau_*} \right)^{2q}\!\!
\left[ \left(dx^1\right)^2 + \left(dx^2\right)^2 + 
\left(dx^3 \right)^2 \right].
\label{eq:2.11}
\end{equation} 
In summary, for  $q=1/3$ we re-obtain the de Sitter expansion, while for  $1/4 < q < 
1/3$, i.e. for $p>1$,  the power-law inflation is realized. Additionally, note that for 
$p <0$, i.e. for $q<0$ or $q>1/3$, we acquire $\dot{H}=-p/t^2 >0$, which corresponds to 
the realization of super-inflation. Finally, for $0 < p \leq 1$, i.e. for $0 < q \leq 
1/4$, accelerated expansion, and thus inflation, is not realized. 

Let us now investigate the observables in the case of power-law inflation. 
As it was shown in \cite{Bamba:2014daa}, it is more convenient to use the e-folding 
number 
$N$ as the independent variable 
(for related considerations, see~\cite{Bamba:2014wda, Mukhanov:2014uwa}). Hence, for every 
function 
$g$ we have that 
$\dot{g}=g'(N) H(N)$, where primes denote derivatives with respect to $N$. Thus, the 
Hubble slow-roll parameters (\ref{epsVbb})-(\ref{xiVbb}) can be re-expressed as
\begin{eqnarray}
\label{epsVbbxx}
&&\!\!\!\!\!\!\!\!\!\!\!\!\!\!
\epsilon_1(N)\equiv-\frac{H'(N)}{H(N)}, 
\\
&&\!\!\!\!\!\!\!\!\!\!\!\!\!\!
\epsilon_2(N) \equiv  \frac{H''(N)}{H'(N)}-\frac{H'(N)}{H(N)},
\label{etaVbbxx}\\
&&\!\!\!\!\!\!\!\!\!\!\!\!\!\!
\epsilon_3(N) \equiv
\left[\frac{H(N)H'(N)}{H''(N)H(N)-H'(N)^2}\right]
\nonumber\\
&&\!\!\!\!
\cdot\left[ \frac{H'''(N)}{H'(N)}- 
\frac{H''(N)^2}{H'(N)^2}-\frac{H''(N)}{H(N)}
+\frac{H'(N)^2}{H(N)^2}
\right].
\label{xiVbbxx}
\end{eqnarray}

Let us now consider the power-law inflation of the form (\ref{apowergener2}) or, 
expressed 
in unimodular terms, of (\ref{eq:2.1200bb}) and (\ref{eq:2.11}). For this case we obtain 
\begin{eqnarray}
&&\epsilon_1=\frac{1}{p}, 
\\
&&\epsilon_2=0, 
\\
&&\epsilon_3=\frac{1}{p}, 
\end{eqnarray}
and thus (\ref{eps111bb})-(\ref{eps444bb}) give 
 \begin{eqnarray}
 &&r \approx\frac{16}{p},
 \label{eps111bbcd}\\
&& n_\mathrm{s} \approx 1-\frac{2}{p} ,
\label{eps2222bbcd} \\
&&\alpha_\mathrm{s} \approx 0 ,
\label{eps333bbcd}\\
 &&n_\mathrm{T} \approx -\frac{2}{p},
\label{eps444bbcd}
\end{eqnarray}
where we remind that $p=q/(1-3q)$. 
Finally, eliminating $p$ between (\ref{eps111bbcd}),(\ref{eps2222bbcd}) we obtain
\begin{eqnarray}
 r=8(1-n_\mathrm{s}).
\end{eqnarray}
Hence, if we take $p=100$, i.e. for $q=0.332$, we acquire $r\approx0.16$
$n_\mathrm{s} \approx 0.98$,
$\alpha_\mathrm{s}=0$, and $n_\mathrm{T} \approx -0.02$, which is in good agreement with 
the Planck results \cite{Planck:2015xua}. However, one can see that the power-law 
inflation is a quite simple scenario and thus the expressions for the observables, 
although in agreement with observations, are simple and one does not have large 
parametric 
freedom to change their values. Hence, in the next subsection we examine a more 
complicated scenario, with improved phenomenological behavior.

\subsection{Starobinsky inflation} 

Let us consider a simple but efficient model, namely the $R^2$ or 
Starobinsky inflation. 
In curvature modified gravity, and in particular in $R^2$ inflation, the action is given 
by 
$S = \int d^4 x \sqrt{-g} \left\{ R + \left[1/\left(6M^2\right)\right] R^2 \right\}$, 
where $M$ is 
a constant with mass dimension. 
In such a scenario the Hubble parameter can be described 
as \cite{Starobinsky:1980te, BC-OO}
\begin{equation}
H = H_* - \frac{M^2}{6} \left(t-t_*\right) \,, 
\label{eq:3.4}
\end{equation}
where $H_*$ is the value of the Hubble function at $t=t_*$. 
Straightforwardly, the scale factor reads as 
$a(t) = a_* \exp \left[H_* \left(t-t_*\right) - \left(M^2/12\right) \left(t-t_*\right)^2 
\right]$. For $t \ll t_*$, that is at the early stages of inflation, 
we approximately acquire 
$a(t) = a_* \exp \left\{ \left[H_* + \left(M^2/6\right) t_* \right]t 
-\left[H_* + \left(M^2/12\right) t_* \right]t_* \right\}$. 
Thus, using the expressions extracted in Section \ref{unimodulars}, and in particular 
(\ref{dtdatu}), in the framework of unimodular gravity and the new time variable we have
\begin{eqnarray} 
&&\!\!\!\!\!\!\!\!\!\!\!\!\!\!\!\!\! \!
\tau = \bar{\tau} 
+ \frac{a_*^3}{3H_* + \left(M^2/2\right) t_*} \exp \left[ 
\left(3H_* + \frac{M^2}{2} t_* \right) t \right.\nonumber\\
&&\left. \ \ \ \ \ \ \ \ \ \ \ \ \ \ \  \ \  \ \  \ \  \ \  \ \  \ \  \ \ 
-\left(3H_* + \frac{M^2}{4} t_* \right) t_* 
\right], 
\label{eq:3.5}\\
&&\!\!\!\!\!\!\!\!\!\!\!\!\!\!\!\!\! \! 
a (\tau) = \left( 3H_* + \frac{M^2}{2} t_* 
\right)^{1/3} 
\left( \tau-\bar{\tau} \right)^{1/3}, 
\label{eq:3.6}
\end{eqnarray} 
where $\bar{\tau}$ is a constant. From (\ref{eq:3.6}) we obtain $\mathcal{H} 
= 1/\left[3\left( \tau-\bar{\tau} \right)\right]$, and therefore, using 
(\ref{Tscalartau}) for the torsion scalar
we acquire $T = -6\left[ H_* + \left(M^2/6\right) t_* \right]^2$. Thus, we deduce that 
in the early stage of inflation $T$ is approximately constant.

The e-folding number $N$ at the inflationary stage is defined as
\begin{equation}
N \equiv \ln \left( \frac{a_\mathrm{f}}{a_\mathrm{i}} \right) 
= - \int_{t_\mathrm{f}}^{t_\mathrm{i}} H (\tilde{t}) d\tilde{t} \,, 
\label{eq:3.3}
\end{equation}
where $a_\mathrm{i} = a(t=t_\mathrm{i})$ is the value of the scale factor $a$ at the 
beginning of inflation $t_\mathrm{i}$, and $a_\mathrm{f} = a(t=t_\mathrm{f})$ is its 
value  at the end 
 of inflation  $t_\mathrm{f}$. Inserting (\ref{eq:3.4}) into 
(\ref{eq:3.3}) we acquire 
\begin{equation}
N = -\left(H_* + \frac{M^2}{6} t_* \right) \left(t_\mathrm{i}-t_\mathrm{f}\right) 
+ \frac{M^2}{12} \left(t_\mathrm{i}^2-t_\mathrm{f}^2\right),
\label{eq:3.8}
\end{equation}
which for positive $t_\mathrm{f}$ can be inverted to express $t_\mathrm{f}$ in terms of 
$N$, namely
\begin{eqnarray}
&&\!\!\!\!\!\!\!\!\!\!\!\!\!\!\!\!\! \! 
t_\mathrm{f}=6\frac{H_*}{M^2}+t_*+M^{-2}\Big\{
\left(6 H_*+M^2 t_*\right)^2
\nonumber\\
&&\
+M^2\left\{t_\mathrm{i}\left[ 
M^2\left(t_\mathrm{i}-2t_*\right)-12 H_*\right]-12N\right\}
\Big\}^{\frac{1}{2}}.
\label{tNrel}
\end{eqnarray}
  
We can now use (\ref{eq:3.4}) in order to 
calculate the Hubble slow-roll parameter $\epsilon_1$ from (\ref{epsVbb}), namely
\begin{eqnarray}
\label{epsVbb200}
&&\!\!\!\!\!\!\!\!\!\!\!\!\!\!
\epsilon_1=\frac{6M^2}{\left[6H_*-M^2\left(t_\mathrm{f}-t_*\right)\right]^2},
\end{eqnarray}
and thus eliminating $t_\mathrm{f}$ in favor of $N$ using (\ref{tNrel}) we obtain 
{\small{
\begin{equation}
\label{epsVbb2002}
\!\!\!\!\!\!
\epsilon_1(N)\!=\!\frac{6M^2}{\left(6 H_*\!+\!M^2 
t_*\right)^2\!+\!M^2\left\{t_\mathrm{i}\left[ 
M^2\left(t_\mathrm{i}-2t_*\right)\!-\!12 H_*\right]\!-\!12N\right\}}.
\end{equation}}}
Similarly, from (\ref{etaVbb}),(\ref{xiVbb}) we obtain
\begin{eqnarray}
&&\!\!\!\!\!\!\!\!\!\!\!\!\!\!
\epsilon_2(N) =2\epsilon_1(N),
\label{etaVbb2}\\
&&\!\!\!\!\!\!\!\!\!\!\!\!\!\!
\epsilon_3(N) = 2\epsilon_1(N).
\label{xiVbb2}
\end{eqnarray}
Inserting these into 
(\ref{eps111bb})-(\ref{eps444bb}) we find 
 \begin{eqnarray}
 r(N) &=&16\, \epsilon_1 (N),
 \label{eps111bb3}\\
 n_\mathrm{s}(N)&=& 1-6\, \epsilon_1 (N) ,
\label{eps2222bb3} \\
\alpha_\mathrm{s}(N) &=& -8\, \epsilon_1^2(N)  ,
\label{eps333bb3}\\
 n_\mathrm{T}(N) &=& -2\, \epsilon_1(N).
\label{eps444bb3}
\end{eqnarray}%
These expressions for the inflationary observables can be in a very good agreement with 
observations \cite{Planck:2015xua}. For instance, taking  $t_\mathrm{i} = 1/H_*$, $t_* = 
3/H_*$,  $H_*/M = 0.04$, and for e-folding number $N=50$, we find  $r\approx0.049$
$n_\mathrm{s} \approx 0.981$,
$\alpha_\mathrm{s}=-7.78 \times 10^{-5}$, and $n_\mathrm{T} \approx -0.0062$, while for 
 $N=60$, we find  $r\approx0.053$
$n_\mathrm{s} \approx 0.98$,
$\alpha_\mathrm{s}=-8.85 \times 10^{-5}$, and $n_\mathrm{T} \approx -0.0067$.

We can eliminate the complicated function $\epsilon_1(N)$ between (\ref{eps111bb3}) and 
(\ref{eps2222bb3}), and between (\ref{eps111bb3}) and (\ref{eps333bb3}),   obtaining 
respectively
\begin{equation}
\label{rnsrel}
r=\frac{8}{3}(1- n_\mathrm{s}),
\end{equation}
and
\begin{equation}
\label{asnsrel}
\alpha_\mathrm{s}=-\frac{2}{9}(1- n_\mathrm{s})^2,
\end{equation}
which prove to be very useful, since they allow us to compare the predictions of our 
scenario with the observational data. In particular, the Planck 
results \cite{Planck:2015xua} 
suggest that $n_{\mathrm{s}} = 0.968 \pm 0.006\, (68\%\,\mathrm{CL})$, 
$r< 0.11\, (95\%\,\mathrm{CL})$, 
and $\alpha_\mathrm{s} = -0.003 \pm 0.007\, (68\%\,\mathrm{CL})$. 
The combined analysis of the BICEP2 and Keck Array data with 
the Planck data shows $r< 0.07\, (95\%\,\mathrm{CL})$ \cite{Array:2015xqh}. As we 
can see, using (\ref{rnsrel}) and (\ref{asnsrel}),  for $n_{\mathrm{s}} \approx 0.97$ we 
obtain $r\approx 0.08$ and  
$\alpha_\mathrm{s} \approx -0.0002 $, which reveals a very good agreement with 
observations.

\subsection{ A specific model: $F(T) =  -T + \alpha_1 (-T)^n+\Lambda$ and $\lambda (T) 
= \alpha_2+\alpha_3 (-T)^m$  } 
\label{specificmodel}

Let us close the investigation of inflationary cosmology in the framework of unimodular 
$F(T)$ gravity, by examining a specific model. As we showed in subsection 
\ref{lagmultform}, in unimodular $F(T)$ gravity inflation may arise from power-law 
forms of the $F(T)$ and Lagrange multiplier $\lambda (T)$ functions. Hence, as a specific 
example we choose 
\begin{eqnarray}
F(T) \Eqn{=}  -T + \alpha_1 (-T)^n+\Lambda \,, 
\label{eq:4.3} \\
\lambda (T) \Eqn{=}\alpha_2+\alpha_3 (-T)^m \,, 
\label{eq:4.4} 
\end{eqnarray}
where $\alpha_1$, $\alpha_2$, $\alpha_3$, $\Lambda$ and 
$n,m$ are constants (the minus sign in front of $T$ is chosen for convenience, since in 
the usual conventions of $f(T)$ formulation, in FRW geometry $T=-6H^2<0$). In the 
following we will focus on the case where $n=2$, $m=1/2$. Inserting the above  ansatzes 
into the first Friedmann equation (\ref{eq:2.17}), using the relation $d\tau = a^3(\tau) 
dt$, and considering also the matter sector to correspond to radiation, i.e 
$\rho_{\mathrm{M}}=\rho_{r0}/a(t)^4$, we obtain the following differential equation:
\begin{equation} 
 \Lambda-\alpha_2-6H(t)^2-\sqrt{6}\alpha_3 H(t)+\frac{\rho_{r0}}{a(t)^4} -108 
\alpha_1 H(t)^4=0,
\label{adiffeq}
\end{equation}
with $H(t)\equiv\dot{a}(t)/a(t)$ and  $\rho_{r0}$ a constant. Choosing small values for 
$\alpha_1$ (such that
$\alpha_1 H(t)^2\ll 1$) and for $\rho_{r0}$ (such that $\rho_{r0}\ll 
4\alpha_2-\alpha_3^2-4\Lambda$) we can extract the solution as
\begin{equation} 
H(t)\approx\frac{\beta_2  }{\beta_3  +2e^{\beta_1t}},
\label{Hgensol}
\end{equation}
where the new constants are defined as
\begin{eqnarray}
&&
\!\!\!\!\!\!\!
\beta_1=\sqrt{\frac{2}{3}}\left(\alpha_3-   \sqrt{
\alpha_3^2+4\Lambda-4\alpha_2}
 \right)
\nonumber\\
&&
\!\!\!\!\!\!\!
\beta_2=
\frac{\rho_{r0}^{-1}}{2\sqrt{6}}\, \Big(\sqrt{\alpha_3^2+4\Lambda-4\alpha_2}-\alpha_3
\Big)\nonumber\\
&& \ \ \ \
\cdot
\exp
\Big\{
4\Big[
(\alpha_3^2\!+\!4\Lambda\!-\!4\alpha_2)\Big(4\alpha_2\!+\!\alpha_3\sqrt{
\alpha_3^2\!+\!4\Lambda\!-\!4\alpha_2}
\Big)
\nonumber\\
&& \ \ \ \ \ \ \ \ \ \ \ \ \ \ \ \,
-\alpha_3^2-4\Lambda 
\Big]C_1
\Big\}
\nonumber\\
&&
\!\!\!\!\!\!\!
\beta_3=
\rho_{r0}^{-1}  \exp\Big\{4\sqrt{\alpha_3^2+4\Lambda-4\alpha_2}
\Big[\alpha_3(1-4\alpha^2+4\Lambda)
\nonumber\\
&& \ \ \ \ \ \ \ \ \  \ 
\ \ \ \ \ +\alpha_3^2
+(4\alpha_2-1)\sqrt{
\alpha_3^2\!+\!4\Lambda\!-\!4\alpha_2}
\Big] C_1\Big\},
\label{const1}
\end{eqnarray}
with $C_1$ is an integration constant.

Let us now use the solution (\ref{Hgensol}) in order to calculate the inflationary 
observables. We start by calculating the e-folding number $N$ defined in (\ref{eq:3.3}), 
obtaining
\begin{equation}
N=\frac{\beta_2}{\beta_3 \beta_1}
\left[\beta_1(t_i-t_f)+
\ln
\left(\frac{\beta_3 +2 e^{\beta_1t_f }    }{\beta_3 
 +2 e^{\beta_1t_i }  }\right)\right],
\label{eq:3.3gensol}
\end{equation}
where $t_\mathrm{i}$ is the beginning of inflation and  $t_\mathrm{f}$  its end. Relation 
(\ref{eq:3.3gensol}) can be easily inverted to give $t_i(N)$. Next, we use 
(\ref{Hgensol})  in oder to  
calculate the Hubble slow-roll parameters   from 
(\ref{epsVbb}),(\ref{etaVbb}),(\ref{xiVbb}), acquiring
\begin{eqnarray}
\label{epsVbb200gen}
&&
\epsilon_1=\frac{2 \beta_1 e^{\beta_1t_i } }{\beta_2}
\\
&&\epsilon_2=
\frac{ \beta_1 \left(\beta_3+2 e^{\beta_1t_i } \right)}{\beta_2}
\\
&&\epsilon_3=\frac{2 \beta_1 e^{\beta_1t_i } }{\beta_2}.
\end{eqnarray}
Inserting these into 
(\ref{eps111bb})-(\ref{eps444bb}) we find 
 \begin{eqnarray}
 &&\!\!\!\!\!\!\!  \!\!\!\!\!\!
 r =
 \frac{32 \beta_1 e^{\beta_1t_i } }{\beta_2},
 \label{eps111bb3gen}\\
 &&\!\!\!\!\!\!\!  \!\!\!\!\!\!
 n_\mathrm{s}= 1-\frac{ 2 \beta_1 \left(\beta_3+4 e^{\beta_1t_i } \right)}{\beta_2}
  ,
\label{eps2222bb3gen} \\
 &&\!\!\!\!\!\!\!  \!\!\!\!\!\!
 \alpha_\mathrm{s} = 
 -\frac{ 8 \beta_1^2  e^{\beta_1t_i }
 \left(\beta_3+2 e^{\beta_1t_i } 
\right)}{\beta_2^2},
\label{eps333bb3gen}\\
 &&\!\!\!\!\!\!\!  \!\!\!\!\!\!
 n_\mathrm{T} =- \frac{4 \beta_1 e^{\beta_1t_i } }{\beta_2}.
\label{eps444bb3gen}
\end{eqnarray}

Hence, we can insert (\ref{eq:3.3gensol}) into the above relations in order to eliminate 
$t_i$ in favor of $N$, obtaining 
 \begin{eqnarray}
 &&\!\!\!\!\!\!\!  \!\!\!\!\!\!\!\!\!\!\!\!
 r(N) =
 \frac{32 \beta_3 \beta_1 e^{\beta_1 \left(\frac{\beta_3}{\beta_2}N +t_f    \right) 
 } }{\beta_2\left[\beta_3-2\left(
e^ {\frac{ \beta_1 \beta_3 } {\beta_2 }N}    
-1
\right)e^{\beta_1 t_f}
\right]},
 \label{eps111bb3genN}\\
 &&\!\!\!\!\!\!\!  \!\!\!\!\!\!\!\!\!\!\!\!
 n_\mathrm{s}(N)
 =
 1-
 \frac{2 \beta_3   \beta_1  
 \left[  \beta_3+2
   e^{\beta_1 t_f}
\left(  1+      e^ {\frac{ \beta_1 \beta_3 } {\beta_2 }N}    \right) 
\right]}
{\beta_2\left[\beta_3-2\left(
e^ {\frac{ \beta_1 \beta_3 } {\beta_2 }N}    
-1
\right)e^{\beta_1 t_f}
\right]}
  ,
\label{eps2222bb3genN} \\
 &&\!\!\!\!\!\!\!  \!\!\!\!\!\!\!\!\!\!\!\!
 \alpha_\mathrm{s}(N) =  
 -
 \frac{8 \beta_3^2   \beta_1^2
 \left( \beta_3 +   2  e^{\beta_1 t_f}   \right)
    e^ { \beta_1 \left(
    \frac{\beta_3}{\beta_2}N +t_f
    \right)      }}
{\beta_2^2
\left[\beta_3-2\left(
e^ {\frac{ \beta_1 \beta_3 } {\beta_2 }N}    
-1
\right)e^{\beta_1 t_f}
\right]},
\label{eps333bb3genN}\\
 &&\!\!\!\!\!\!\!  \!\!\!\!\!\!\!\!\!\!\!\!
 n_\mathrm{T}(N) = 
 -
 \frac{4 \beta_3 \beta_1 e^{\beta_1 \left(\frac{\beta_3}{\beta_2}N +t_f    \right)  
} }{\beta_2\left[\beta_3-2\left(
e^ {\frac{ \beta_1 \beta_3 } {\beta_2 }N}    
-1
\right)e^{\beta_1 t_f}
\right]}
 .
\label{eps444bb3genN}
\end{eqnarray}
Finally, we can insert (\ref{eps2222bb3genN}) into (\ref{eps111bb3genN}) and  
(\ref{eps333bb3genN}),(\ref{eps444bb3genN}),
in order to  eliminate $N$ and 
$t_f$ in favor of $ n_\mathrm{s}$, resulting to
  \begin{eqnarray}
 &&\!\!\!\!\!\!\!  \!\!\!\!\!\!\!\!\!
 r =4(1   -n_\mathrm{s} )
 -\frac{8 \beta_3  \beta_1 }{\beta_2}
 ,
 \label{eps111bb3genNb00}\\
 &&\!\!\!\!\!\!\!  \!\!\!\!\!\!\!\!\!
 \alpha_\mathrm{s} =-
 \frac{1}{4} 
  ( n_\mathrm{s}-1)^2+   \frac{  \beta_3^2  \beta_1^2    }{ \beta_2^2} 
,
\label{eps333bb3genNb00}\\
 &&\!\!\!\!\!\!\!  \!\!\!\!\!\!\!\!\!
 n_\mathrm{T} = -\frac{1}{2}(1   -n_\mathrm{s} )
 +\frac{ \beta_3  \beta_1 }{\beta_2},
\label{eps444bb3genNb00}
\end{eqnarray}
or, restoring the original parameters using (\ref{const1}), we obtain
{\small{
  \begin{eqnarray}
 &&\!\!\!\!\!\!\!\!\!\!\!\!\!\!\!\!\!\!\!\!\!
 r =4(1   -n_\mathrm{s} )\nonumber\\
 &&\!\!\!\!\!\!\!\!\!\!\!\!\!\!  \!\!\!\!\!\!\!\!
 +32\exp\Bigg\{\!
 \frac{4\left[\alpha_3^3\!
 +\!4\alpha_3\Lambda\!+\!4\alpha_2\left(
\sqrt{\alpha_3^2\!+\!4\Lambda\!-\!4\alpha_2}\!-\!\alpha_3
 \right)
 \right]\!C_1}
 {\sqrt{\alpha_3^2+4\Lambda-4\alpha_2}}\!
\Bigg\} \!
 ,
 \label{eps111bb3genNb}\\
 &&\!\!\!\!\!\!\!\!\!\!\!\!\!\!\!\!\!\!\!\!\!
 \alpha_\mathrm{s} =  -
 \frac{1}{4} 
  ( n_\mathrm{s}-1)^2\nonumber\\
  &&\!\!\!\!\!\!\!\!\!\!\!\!\!\!  \!\!\!\!\!\!\!\!
  +
 16\exp\Bigg\{\!
 \frac{4\left[\alpha_3^3\!
 +\!4\alpha_3\Lambda\!+\!4\alpha_2\left(
\sqrt{\alpha_3^2\!+\!4\Lambda\!-\!4\alpha_2}\!-\!\alpha_3
 \right)
 \right]\!C_1}
 {\sqrt{\alpha_3^2+4\Lambda-4\alpha_2}}\!
\Bigg\}  \!
,
\label{eps333bb3genNb}\\
 &&\!\!\!\!\!\!\!\!\!\!\!\!\!\!\!\!\!\!\!\!\!
 n_\mathrm{T} = -\frac{1}{2}(1   -n_\mathrm{s} )\nonumber\\
 &&\!\!\!\!\!\!\!\!\!\!\!\!\!\!  \!\!\!\!\!\!\!\!
-4\exp\Bigg\{\!
 \frac{4\left[\alpha_3^3\!
 +\!4\alpha_3\Lambda\!+\!4\alpha_2\left(
\sqrt{\alpha_3^2\!+\!4\Lambda\!-\!4\alpha_2}\!-\!\alpha_3
 \right)
 \right]\!C_1}
 {\sqrt{\alpha_3^2+4\Lambda-4\alpha_2}}\!
\Bigg\} .
\label{eps444bb3genNb}
\end{eqnarray}}}

\begin{figure}[ht]
\centering
\includegraphics[scale=.50]{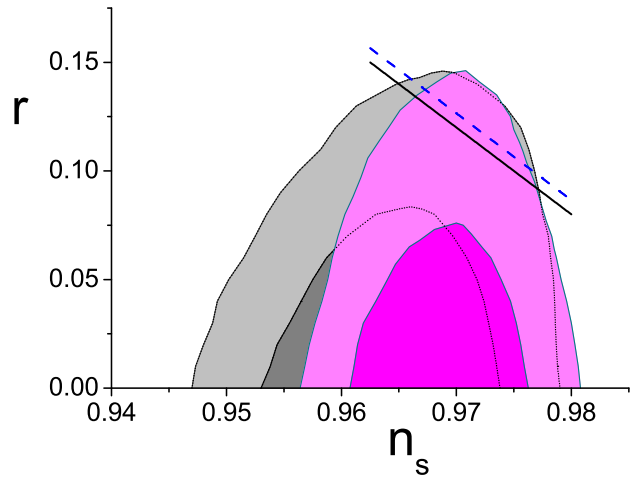}
\caption{{\it{ 1$\sigma$ (magenta) and 2$\sigma$ (light magenta) contours for Planck 2015
results ($TT+lowP+lensing+BAO+JLA+H_0$)  \cite{Ade:2015xua}, and 1$\sigma$ (grey) and
2$\sigma$ (light grey) contours for Planck 2013 results  ($Planck+WP+BAO$)
\cite{Planck:2013jfk} (note that the  1$\sigma$ region of Planck 2013 results  is behind
the Planck 2015 results, hence we mark its boundary by a dotted curve), on 
$n_{\mathrm{s}}-r$ plane.
Additionally, we present the predictions of the specific scenario 
(\ref{eq:4.3})-(\ref{eq:4.4}) of
inflation in unimodular $F(T)$ gravity, with  $n=2$, $m=1/2$, and for  $\alpha_2=0.5$, 
$\alpha_3=2$,  $\Lambda=0.1$, $C_1=-1$ (black - solid curve),    
and for  $\alpha_2=1$, 
$\alpha_3=0.5$,  $\Lambda=1$, $C_1=-0.5$    (blue - dashed 
curve), in units where $2{\kappa}^2 = 1$.
}}}
\label{rns}
\end{figure}

\begin{figure}[ht]
\centering
\includegraphics[scale=.47]{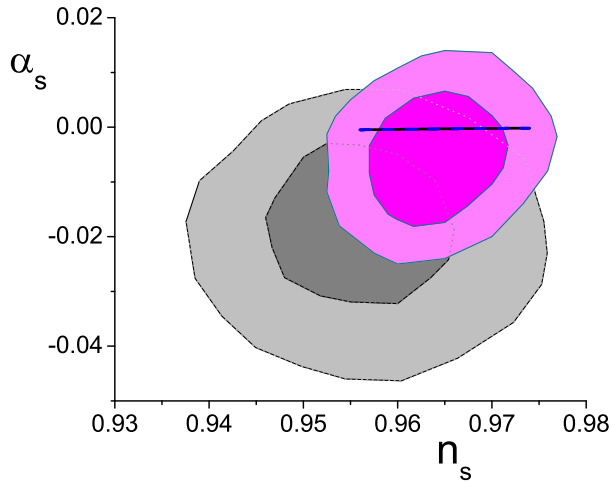}
\caption{{\it{ 1$\sigma$ (magenta) and 2$\sigma$ (light magenta) contours for Planck 2015
results ($TT,TE,EE+lowP$)  \cite{Ade:2015xua}, and 1$\sigma$ (grey) and
2$\sigma$ (light grey) contours for   Planck 2013 results  ($\Lambda
CDM+running+tensors$)
\cite{Planck:2013jfk}, on $n_{\mathrm{s}}-\alpha_\mathrm{s}$ plane. Additionally, we 
present the predictions of the specific scenario 
(\ref{eq:4.3})-(\ref{eq:4.4}) of
inflation in unimodular $F(T)$ gravity,  with  $n=2$, $m=1/2$, and for  $\alpha_2=0.5$, 
$\alpha_3=2$,  $\Lambda=0.1$, $C_1=-1$ (black - solid curve),    
and for  $\alpha_2=1$, 
$\alpha_3=0.5$,  $\Lambda=1$, $C_1=-0.5$    (blue - dashed 
curve), in units where $2{\kappa}^2 = 1$. The two curves are indistinguishable in 
the resolution scale of the figure.   }}}
\label{asns}
\end{figure}

Relations (\ref{eps111bb3genNb})-(\ref{eps444bb3genNb}) prove to be very useful, since 
they allow us to compare the 
predictions of our scenario with the observational data. In particular, in
Fig.~\ref{rns} we use (\ref{eps111bb3genNb}) and we present the estimated 
tensor-to-scalar ratio of the specific scenario 
(\ref{eq:4.3})-(\ref{eq:4.4}) of
inflation in unimodular $F(T)$ gravity, for two cases, on top of 
the 
1$\sigma$ and 
2$\sigma$ contours of
the Planck 2013 results \cite{Planck:2013jfk} as well as of the  Planck 2015 results
\cite{Ade:2015xua}. Additionally, in Fig. \ref{asns} we use (\ref{eps333bb3genNb}) and we 
show the predictions of 
our scenario for the running  spectral index $\alpha_\mathrm{s}$  on top of the 
1$\sigma$ and 2$\sigma$ contours of the Planck 2013
results \cite{Planck:2013jfk} as well as of the  Planck 2015 results
\cite{Ade:2015xua}. The  agreement with observations is inside the 2$\sigma$ region for 
the tensor-to-scalar ratio and it is very satisfactory for the running  spectral index. 
We mention that the agreement with observations is obtained through the unimodular 
construction, since taking $\lambda$ to zero, i.e. for $\alpha_2,\alpha_3\rightarrow0$, 
one arrives to unacceptable deviations. 

\section{Graceful exit from the inflationary stage} 
\label{grexit}

In this section we are interested in investigating the graceful exit from inflation 
in unimodular $F(T)$ gravity. In general, if the de Sitter solution describing the 
inflationary stage is unstable, then the graceful exit from 
inflation can be realized. When the de Sitter inflation occurs
the Hubble parameter is described by 
$H = H_\mathrm{inf}$, where $H_\mathrm{inf}$ is a positive constant. 
In order to examine the instability of the de Sitter solution, in general one  considers 
the perturbations of 
the Hubble parameter, which can be expressed as 
\begin{eqnarray}
H \Eqn{=} H_\mathrm{inf} + H_\mathrm{inf} \delta(t) \,, 
\label{eq:4.1} \\
\delta(t) \Eqn{\equiv} \delta_0 e^{\beta t} \,, 
\label{eq:4.2} 
\end{eqnarray}
with $\delta_0$ and $\beta$ constants. In   (\ref{eq:4.1})   
 the   term $H_\mathrm{inf} \delta(t)$ describes 
 the perturbations from the de Sitter solution, i.e.  
the constant part $H_\mathrm{inf}$ of the Hubble parameter, and thus we assume
$\left| \delta(t) \right| \ll 1$. Note that in this way we can quantify  
the instability of the de Sitter solution. 
If we obtain a positive solution of $\beta$ then the value of $\left| \delta(t) \right|$ 
increases in time during inflation, and thus   
the de Sitter solution is unstable.  Consequently, the inflationary stage  
ends successfully and the universe can enter the reheating stage and its standard thermal 
history. 

Let us focus on the specific scenario analyzed in subsection \ref{specificmodel}. In this 
case, the solution for the Hubble function (\ref{Hgensol}), in the $\left| \delta(t) 
\right| \ll 1$ regime, can take the form  (\ref{eq:4.1}) with $ 
H_\mathrm{inf}=\beta_2/\beta_3$, $\delta_0=-2/\beta_3$ and $\beta=\beta_1$. Hence, one can 
easily deduce that for $\beta_1>0$ we have a graceful exit from the inflationary regime.

We close this section by making some comments on the difference   between the 
present scenario of inflation in unimodular $F(T)$ gravity and 
$R^2$ inflation (Starobinsky inflation), focusing on the graceful exit. 
In particular, in $R^2$ inflation, for the flat FLRW geometry, 
by differentiating an equation derived from 
the gravitational field equations in the absence of matter \cite{R-D-M}, in terms of $t$, 
one finds 
%
\begin{equation} 
\dddot{R} + 3\dot{H}\dot{R} +3H\ddot{R} +M^2 \dot{R} =0 \,,
\label{eq:4.8}
\end{equation}
where $R = 6\left(H^2 +\dot{H}\right)$. 
Inserting (\ref{eq:4.1}), with $\delta(t) = e^{ \tilde{\beta} t}$, 
into Eq.
~(\ref{eq:4.8}) and keeping first-order terms in $\delta(t)$, we eventually  acquire 
\begin{equation}
  \tilde{\beta} 
\left[ \tilde{\beta}^3 + 10H_\mathrm{inf} \tilde{\beta}^2 + 
\left( 24H_\mathrm{inf}^2 + M^2 \right) \tilde{\beta} +4H_\mathrm{inf} M^2  \right]   = 
0.
\label{eq:4.9}
\end{equation}
It is seen from Eq.~(\ref{eq:4.9}) that even if there   exists
a real solution for $\tilde{\beta}$, it would be a non-positive solution, which 
 implies that through the perturbative analysis  the instability of the de Sitter solution 
does not appear. In summary,   the de Sitter solution for $R^2$ inflation would be more 
stable than that for inflation in unimodular $F(T)$ gravity. This difference between 
$R^2$ inflation and  inflation in unimodular $F(T)$ gravity acts as an advantage for the 
latter.

\section{Conclusions}
\label{Conclusions}

In the present paper, we have investigated the inflationary realization in the context of 
unimodular $F(T)$ gravity. The action of the theory is based on the $F(T)$ modification 
of teleparallel gravity, in which one imposes the unimodular condition through the 
use of Lagrange multipliers. Hence, we have developed the general reconstruction 
procedure of the $F(T)$ form that can give rise to a given scale-factor evolution.

Having presented the general machinery, we have applied it in the inflationary regime. In 
particular, we have extracted the Hubble slow-roll parameters that allow us to calculate 
various inflation-related observables, such as the scalar spectral index and its running, 
the tensor-to-scalar ratio, and the tensor spectral index. Then, we examined the 
particular cases of de Sitter and power-law inflation, of Starobinsky inflation, as well 
as inflation in a specific model of unimodular $F(T)$ gravity. As we showed, in all cases 
the predictions of our scenarios are in a very good agreement with observational data 
from Planck probe.

Apart from the very satisfactory agreement with observations, the scenario of inflation 
in unimodular $F(T)$ gravity has the additional advantage that it always allows for a 
graceful exit for specific regions of the model parameters, as it can be seen by 
examining the instability of the de Sitter phase. This is in contrast with inflationary 
realizations in curvature-based modified gravity, such as the Starobinsky inflation, where 
a graceful exit is not guaranteed. The above features make inflation in unimodular $F(T)$ 
gravity a successful candidate for the description of the early universe.




\section*{Acknowledgments}

This work was partially supported by the JSPS Grant-in-Aid for 
Young Scientists (B) \# 25800136 and 
the research-funds provided by Fukushima University (K.B.), and 
MINECO (Spain) project FIS2013-44881 (S.D.O.).




\end{document}